\documentclass[a4paper]{PoS}
\usepackage{subfigure}
\usepackage{units}
\usepackage{amsmath}
\usepackage{setspace}
\usepackage{multirow}
\usepackage{lipsum}

\let\OLDthebibliography\thebibliography
\renewcommand\thebibliography[1]{
  \OLDthebibliography{#1}
  \setlength{\parskip}{0pt}
  \setlength{\itemsep}{2.2pt plus 1ex}
}

\title{NEUT development for T2K and relevance of updated 2p2h models}
\ShortTitle{NEUT development for T2K}

\author{\speaker{Callum Wilkinson}
       \thanks{On behalf of the T2K collaboration}\\
       The University of Sheffield, Department of Physics and Astronomy, Sheffield, S3 7RH, England\\
       E-mail: \email{callum.wilkinson@sheffield.ac.uk}}

\abstract{The MiniBooNE large axial-mass anomaly has motivated the development of new theoretical Charged Current Quasi-Elastic (CCQE) cross-section models in recent years. These proceedings review the development of the neutrino simulation generator NEUT to incorporate these more sophisticated CCQE models, including multi-nucleon interaction (2p2h) effects. The fit results on the MINER$\nu$A and MiniBoone data are used to tune neutrino interaction models in NEUT and develop a default cross-section model for T2K.}

\FullConference{16th International Workshop on Neutrino Factories and Future Neutrino Beam Facilities\\
                 25 -30 August, 2014\\
                 University of Glasgow, United Kingdom}

\begin{document}

\section{Introduction}
The relativistic Fermi gas (RFG) model of Llewellyn-Smith~\cite{llewellyn-smith} has been used in Monte Carlo (MC) generators to describe Charged-Current Quasi-Elastic (CCQE) interactions for the last 40 years due to its simplicity. The only free model parameter unconstrained by electron-scattering data is the axial-mass, $M_{\mathrm{A}}$, which is well constrained by fits to neutrino-nucleon scattering and pion production data as $M_\mathrm{A} = 1.014 \pm 0.014$~\unit{GeV}. Recent measurements of $M_{\mathrm{A}}$ by experiments using a heavy nuclear target have found inconsistent results compared with the global dataset~\cite{K2K_2006, mb-ccqe-2010, minos_2014, t2k_2014}. In particular, MiniBooNE found a value of $M_\mathrm{A} = 1.35 \pm 0.17$~\unit{GeV} with a shape-only fit to their data~\cite{mb-ccqe-2010}, which has led to the name ``MiniBooNE large axial-mass anomaly''.

There has been a great deal of recent theoretical activity on nuclear effects to explain the discrepancy (for a recent review, see reference~\cite{hayato_2014}). There has been a corresponding push within the T2K Neutrino Interactions Working Group to develop NEUT~\cite{hayato:neut}, T2K's primary interaction generator, to keep up to date with these developments. This report overviews the model developments in NEUT in Section~\ref{sec:models}. There is a brief overview of where external data fits enter into the general T2K analysis framework in Section~\ref{sec:analysis}. A fitting package has been developed to fit NEUT MC to published cross-section data. This package is used to fit all of the available CCQE data to the new CCQE models available in NEUT as described in Section~\ref{sec:fit}, and the results are used to select the default T2K MC model. The Parameter Goodness of Fit (PGoF) test~\cite{pgof-2003} is used to test the consistency of the datasets within each model. An error inflation procedure based on the PGoF test is described and the final results are presented. The best fit parameter values and inflated parameter errors will be used as inputs to T2K oscillation fits, and for near detector cross-section measurements. The work is summarised in Section~\ref{sec:summary}.

\section{NEUT CCQE model developments}\label{sec:models}
NEUT is the primary interaction generator for the T2K experiment, and developments to its interaction model must support all of the target materials used in T2K. The near detector is predominantly composed of hydrocarbon and water targets, and the far detector, Super-Kamiokande, is a water target. Additional chemical elements present in the near detector, but not used as a primary target for most analyses are iron, lead, brass and argon, which must also be simulated.

Two additional nuclear models have been implemented in NEUT to provide alternatives to the RFG model. The Benhar Spectral Function (SF) model~\cite{sf} is a more realistic two-dimensional description of the initial state nucleon within the nucleus in terms of its momentum and removal energy. This SF model includes the effect of short-range correlations within the nucleus, accounting for $\sim$20\% of the total cross-section. Note that the impulse approximation is still assumed: the interaction is always with a single nucleon. The effective spectral function model of Bodek {\it et al.}~\cite{eff-sf} has also been implemented in NEUT, but it is not ready to be a default model for T2K, so this is not discussed further in this work.

Two nuclear effects which are not part of the RFG model have also been included. The Random Phase Approximation (RPA) is a nuclear screening effect due to long range nucleon-nucleon correlations~\cite{nieves}. Two RPA models, relativistic and non-relativistic, are available from the same authors. These are different treatments for the quenching of the RPA effect at high four momentum transfer, $Q^{2}$. The NEUT implementation of both of these RPA models is dependent on $Q^2$ and $E_{\nu}$. Additionally, a meson exchange current model by Nieves (MEC)~\cite{nieves, nievesExtension} has been implemented (full details are given in~\cite{peter_nuint}). MEC, or 2p2h models, go beyond the impulse approximation, where nucleons within the nucleus are treated as quasi-free. Note that only the lepton kinematics are available for the Nieves MEC model, T2K does not have access to the nucleon predictions. An effective multi-nucleon ejection model~\cite{multinucleon} is used to simulate the outgoing nucleons, along with the NEUT FSI cascade model. As we do not currently have a fully consistent description of the hadronic system, we only fit to lepton kinematics in this work.

With these ingredients, there are two model combinations which can be used as the T2K default MC model:
\vspace{-\topsep}
\begin{enumerate}
  \setlength{\itemsep}{1pt}
  \setlength{\parskip}{0pt}
  \setlength{\parsep}{0pt}
  \item SF+MEC
  \item RFG+RPA+MEC
\end{enumerate}
\vspace{-\topsep}
Note that the SF+MEC model is incomplete without RPA. As we do not currently have access to an appropriate RPA calculation for the SF model, this incomplete model is the best we have available. The BBBA05 form factors~\cite{bbba05} are used consistently for both models throughout this work.

\section{T2K analysis structure}\label{sec:analysis}
\begin{figure}[htb]
    \centering
    \includegraphics[width=0.6\textwidth]{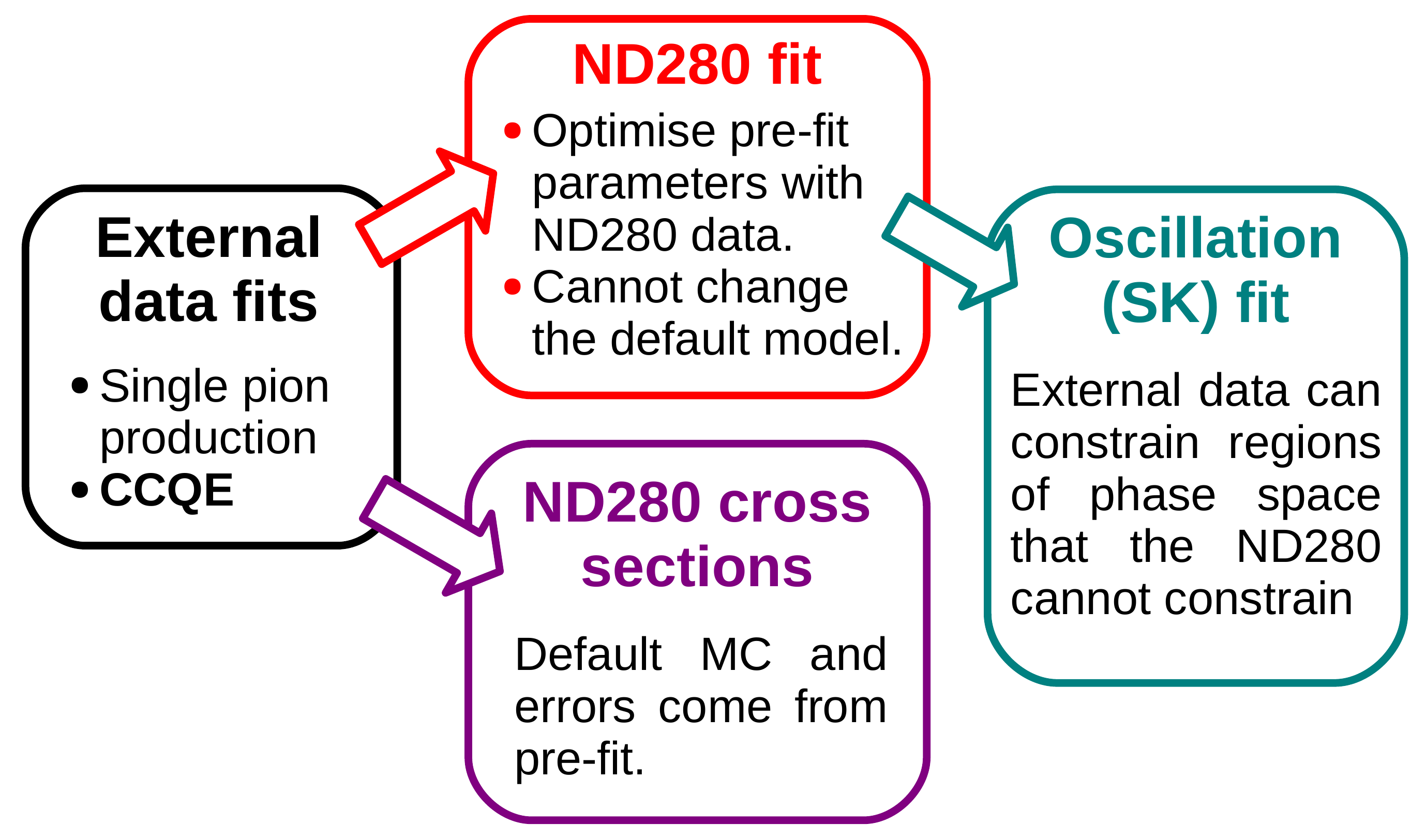}
    \caption[SILLY LATEX WORKAROUND]{A flow diagram illustrating how external data fits are used an an input into various T2K analyses.}
    \label{fig:t2k_analysis_structure}
  \end{figure}

External data fits form an integral part of the T2K analysis structure, as is illustrated by the flow diagram in Figure~\ref{fig:t2k_analysis_structure}. Fits to CCQE and resonant pion-production channels are performed independently and provide inputs to the near detector (ND280) fit and cross-section measurements. The ND280 fit constrains all of the flux and cross-section parameters which are then propagated to the far detector for neutrino oscillation measurements~\cite{t2k_2013}.
\section{CCQE fits to external data}\label{sec:fit}
Four datasets are used for the CCQE fits presented in these proceedings. These datasets are: the MiniBooNE neutrino~\cite{mb-ccqe-2010} and antineutrino~\cite{mb-ccqe-antinu-2013} results on a CH$_2$ target, which are double-differential in cosine of the muon angle, $\cos\theta_{\mu}$, and the muon kinetic energy, $T_{\mu}$; and the MINER$\nu$A neutrino~\cite{minerva-nu-ccqe} and antineutrino~\cite{minerva-antinu-ccqe} results on a CH target, which are differential in reconstructed four-momentum transfer, $Q^{2}_{\mathrm{QE}}$, defined in Equation~\ref{eq:q2qe}:
\begin{align}
  \small
  E_{\nu}^{\mathrm{QE}} &= \frac{2M_{n}'E_{\mu} - (M_{n}'^{2} + m^{2}_{\mu} - M_{p}^{2})}{2(M'_{n} - E_{\mu} + \sqrt{E^{2}_{\mu} - m^{2}_{\mu}})\cos\theta_{\mu}},\notag\\
  Q^{2}_{\mathrm{QE}} &= -m_{\mu} + 2E_{\nu}^{\mathrm{QE}}(E_{\mu} - \sqrt{E_{\mu}^{2} - m_{\mu}^{2}}\cos\theta_{\mu}),
  \label{eq:q2qe}
\end{align}
\noindent where $E_{\mu}$ is the muon energy; $M_{n}$, $M_{p}$ and $m_{\mu}$ are the masses of the neutron, proton and muon. And where $M' = M_{n} - V$, with V as the binding energy of carbon assumed in the analysis. For the MINER$\nu$A neutrino (antineutrino) dataset, $V$ = \unit[34]{MeV} ($V$ = \unit[30]{MeV}). The restricted phase-space MINER$\nu$A results, where $\theta_{\mu} \leq 20^{\circ}$, were used to reduce model dependence introduced by corrected to unsampled regions of phase-space.

An MC prediction for an arbitrary set of model parameters, $\vec{\mathbf{x}}$, is produced by reweighting a NEUT sample generated using the relevant experimental flux and target. When fitting to the data, MINUIT is used to minimise the $\chi^{2}$ statistic defined in Equation~\ref{eq:chi2}:
\begin{align}
  \small
            \chi^{2}(\vec{\mathbf{x}}) &= \left\lbrack\sum^{N}_{k=0} \left(\frac{\nu_{k}^{DATA}-\lambda_{\alpha}^{-1}\nu_{k}^{MC}(\vec{\mathbf{x}})}{\sigma_{k}} \right)^{2} 
            + \left(\frac{\lambda_{\alpha} - 1}{\varepsilon_{\alpha}} \right)^{2}\right \rbrack \rightarrow \mathrm{MiniBooNE~\nu} \notag \\\notag
            &+ \left\lbrack\sum^{M}_{l=0} \left(\frac{\nu_{l}^{DATA}-\lambda_{\beta}^{-1}\nu_{l}^{MC}(\vec{\mathbf{x}})}{\sigma_{l}} \right)^{2} 
            + \left(\frac{\lambda_{\beta} - 1}{\varepsilon_{\beta}} \right)^{2}\right\rbrack \rightarrow \mathrm{MiniBooNE~\bar{\nu}}\\
            &+ \left\lbrack\sum_{i=0}^{16} \sum_{j=0}^{16} \left(\nu_{i}^{DATA} - \nu_{i}^{MC}(\vec{\mathbf{x}})\right)V_{ij}^{-1}\left(\nu_{j}^{DATA} - \nu_{j}^{MC}(\vec{\mathbf{x}})\right)\right\rbrack \rightarrow \mathrm{MINER}\nu\mathrm{A}
\label{eq:chi2}
\end{align}
\noindent where $\vec{\mathbf{x}}$ are the model parameters and $V_{ij}$ is the 16$\times$16 covariance matrix provided by MINER$\nu$A which includes cross-correlations between the neutrino (8 bins) and antineutrino (8 bins) datasets. $\lambda_{\alpha}$ and $\lambda_{\beta}$ are the normalisation parameters for MiniBooNE neutrino and antineutrino respectively, with published normalisation uncertainties of $\varepsilon_{\alpha}$ (10.7\%) and $\varepsilon_{\beta}$ (13.0\%). Fits can be performed to subsets of the data by only including relevant terms in Equation~\ref{eq:chi2}. When fitting to a single MINER$\nu$A dataset, the relevant 8$\times$8 portion of $V_{ij}$ is used, cross-correlations between the neutrino and antineutrino samples are neglected. Note that the MiniBooNE data has no covariance matrix, no bin to bin correlations were published.

The parameters which can be reweighted in NEUT, and which are investigated as fit parameters in this work are:
\vspace{-\topsep}
\begin{itemize}
  \setlength{\itemsep}{1pt}
  \setlength{\parskip}{0pt}
  \setlength{\parsep}{0pt}
  \item MEC normalisation, as a percentage of the nominal Nieves model.
  \item The axial-mass, $M_{\mathrm{A}}$.
  \item The Fermi momentum, $p_{\mathrm{F}}$, which is different for the SF and RFG models.
  \item The overall normalisation of the CCQE cross-section.
\end{itemize}
\vspace{-\topsep}
\subsection{Combined fit}
The best fit $\chi^{2}$ and parameter values from a combined fit to all four datasets is shown in Table~\ref{tab:best_fit} for the SF+MEC and both RFG+RPA+MEC models. The relativistic RPA model was favoured over the non-relativistic model, so only the former is considered further in these proceedings. The best fit parameter values for the SF+MEC model have an inflated axial mass, are at the limit of no MEC, indicating considerable tensions between the parameters favoured by different datasets. None of the datasets favour similar parameters when fit with the SF+MEC model individually, which should be interpreted as evidence that the SF+MEC model cannot fit all of the datasets in a consistent way, despite the comparable $\chi^{2}$ obtained for the SF+MEC and relativistic RFG+RPA+MEC models.

\begin{figure}[htb]
    \centering
    \subfigcapskip=-7pt
    \subfigure[Neutrino]     {\includegraphics[width=0.48\textwidth]{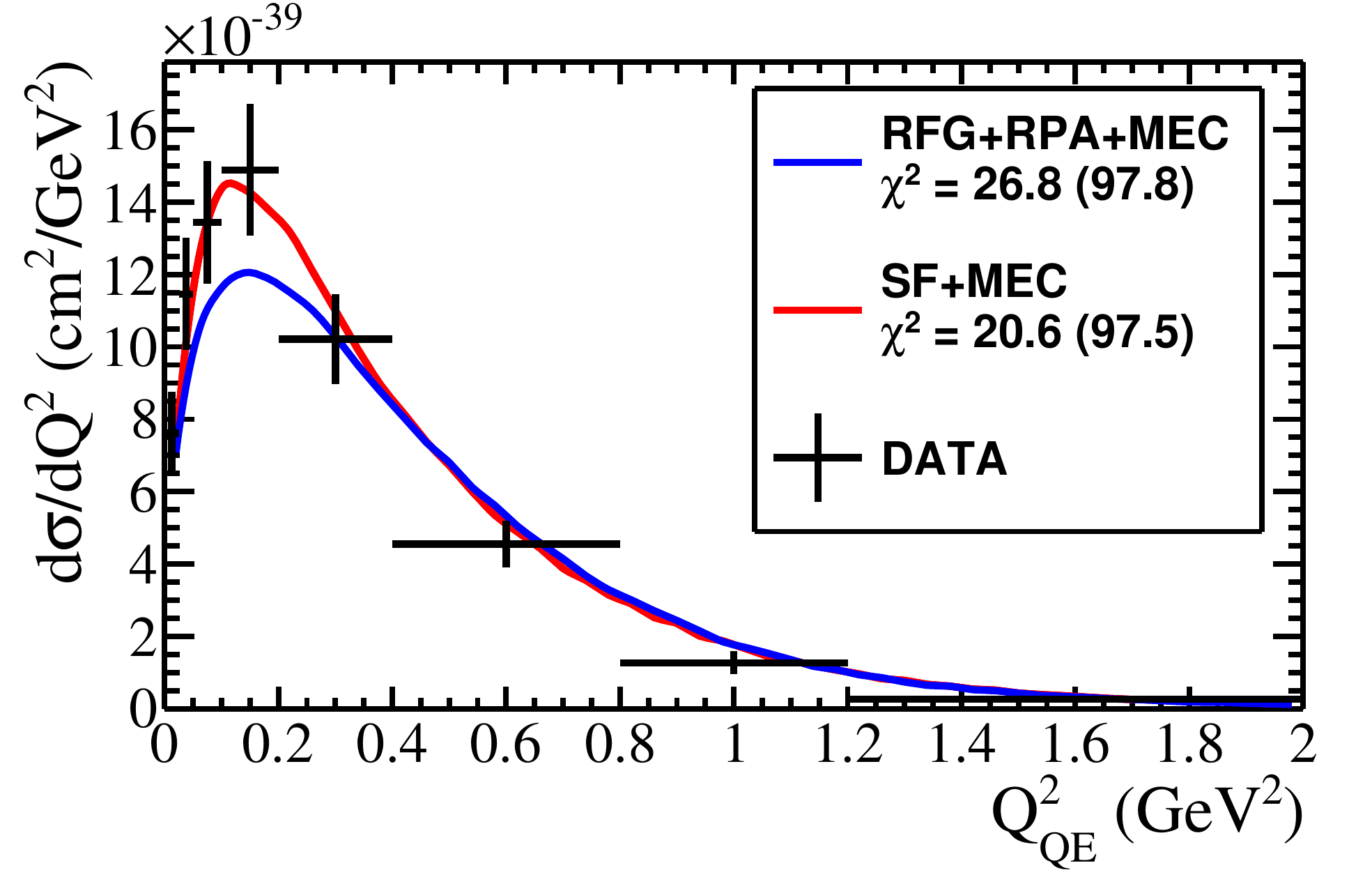}}
    \subfigure[Antineutrino] {\includegraphics[width=0.48\textwidth]{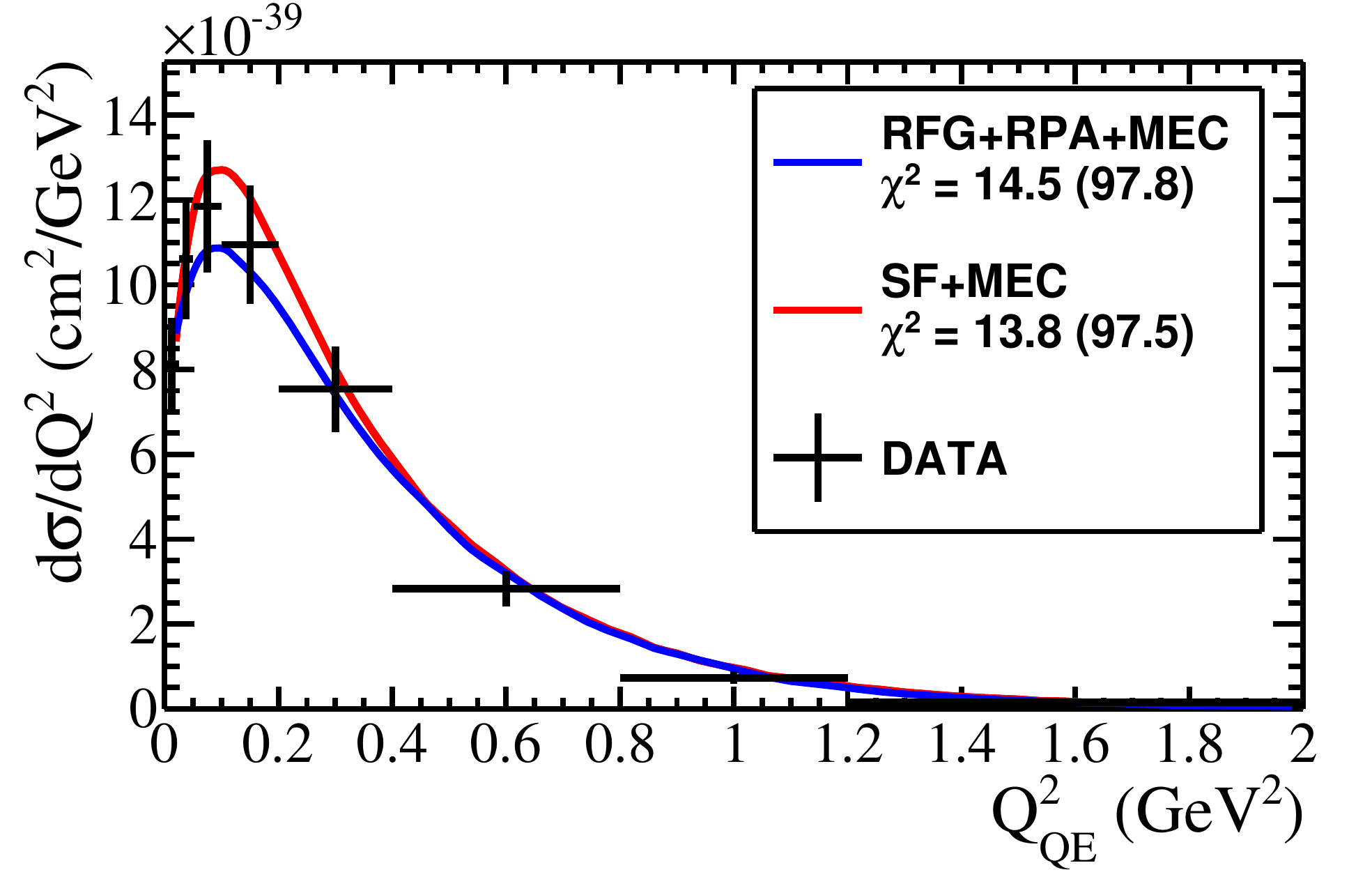}}
    \caption[SILLY LATEX WORKAROUND]{Comparison between the best fit distributions from the combined fits detailed in Table~\ref{tab:best_fit} and the MINER$\nu$A datasets used in the fit. The $\chi^{2}$ value in the legend is the contribution from the dataset shown in the histogram, with the total $\chi^{2}_{min}$ from all four datasets shown in parentheses.}
    \label{fig:best_fit_minerva}\vspace{-12pt}
  \end{figure}

\begin{figure}[htb]
    \centering
    \subfigure[Neutrino]     {\includegraphics[width=0.7\textwidth]{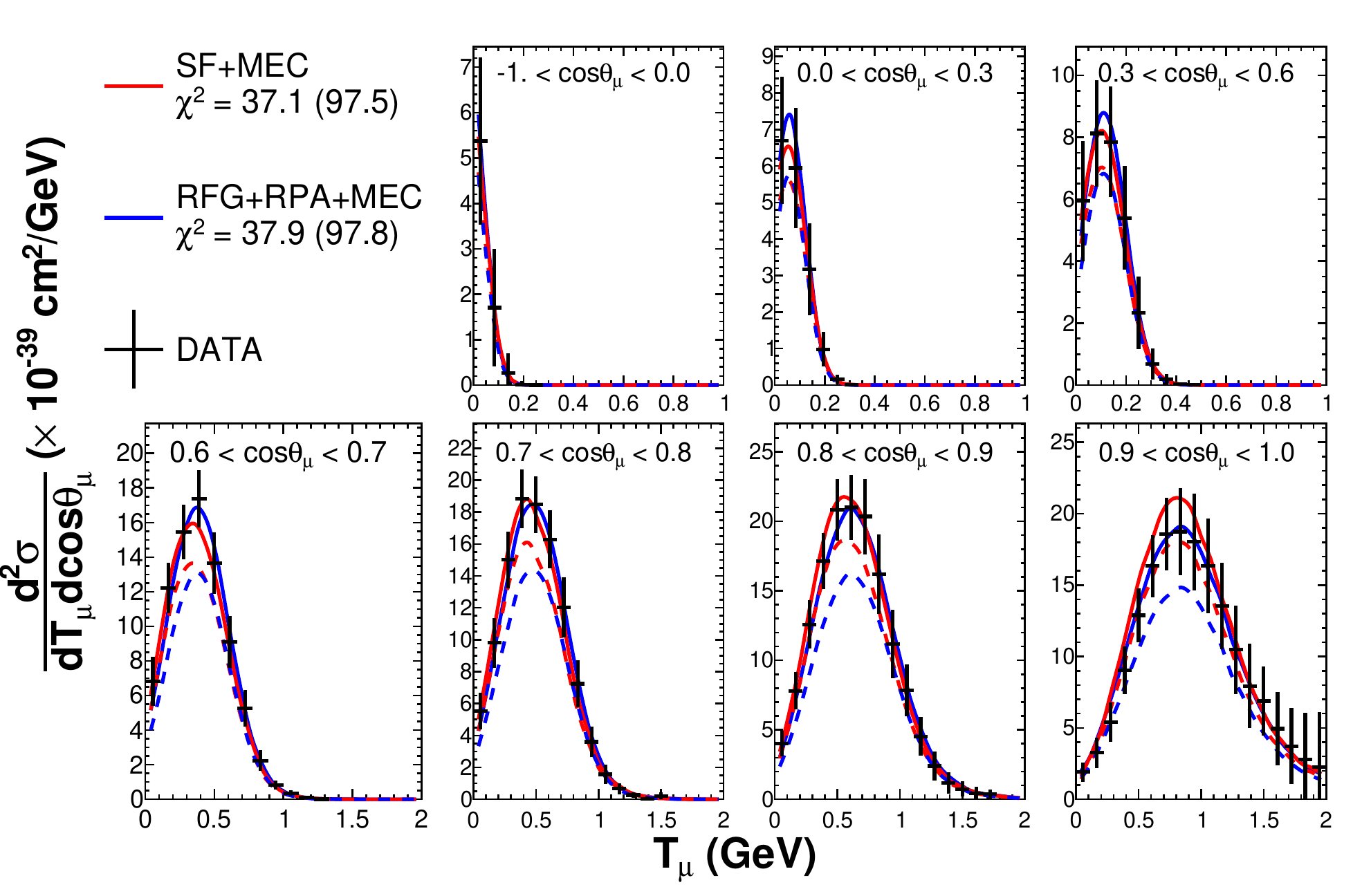}}
    \subfigure[Antineutrino] {\includegraphics[width=0.7\textwidth]{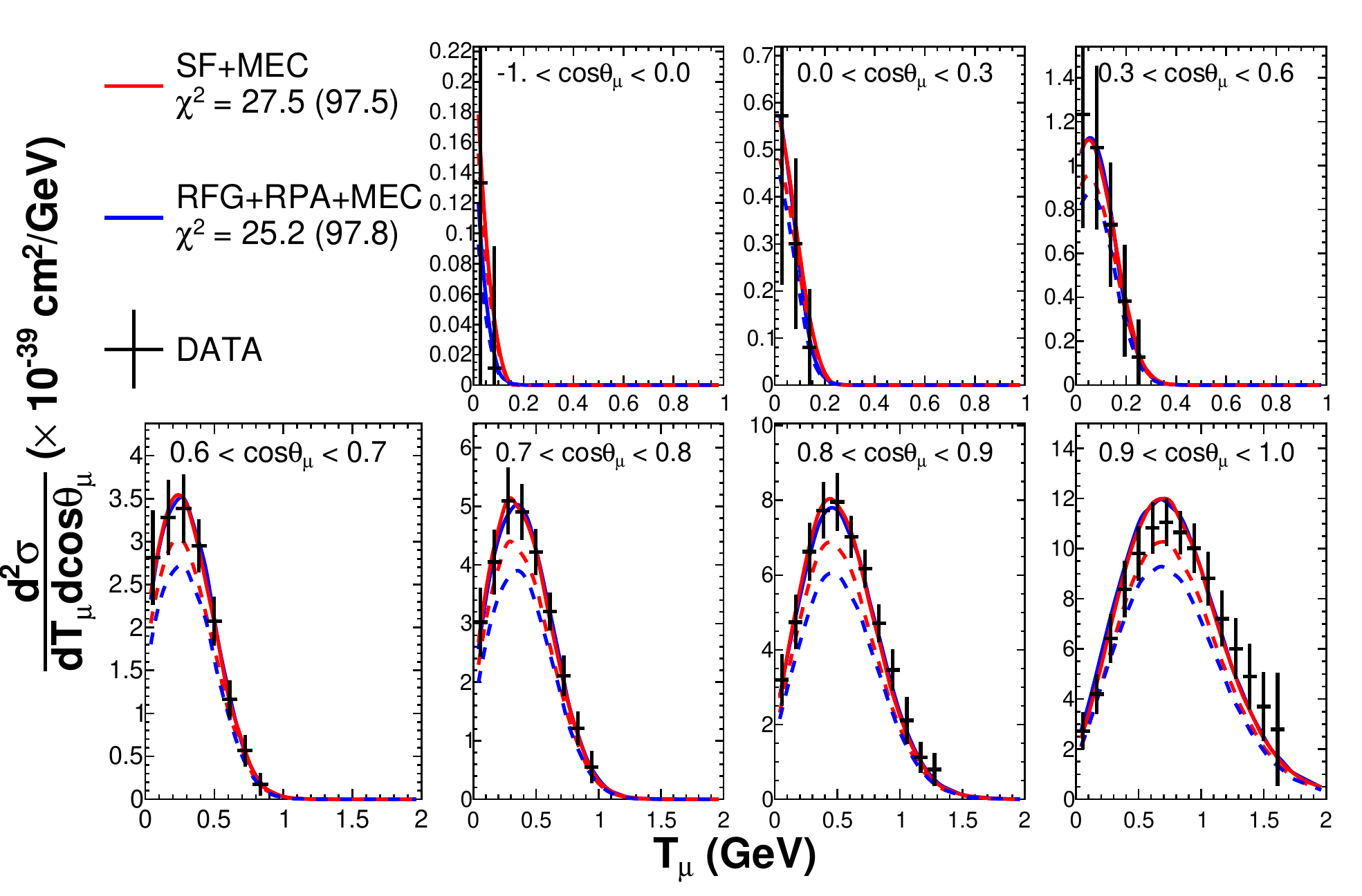}}
    \caption[SILLY LATEX WORKAROUND]{Comparison between the best fit distributions from the combined fits detailed in Table~\ref{tab:best_fit} and the MiniBooNE double-differential datasets used in the fit. The $\chi^{2}$ value in the legend is the contribution from the dataset shown in the histogram, and the total $\chi^{2}_{min}$ from all four datasets shown in parentheses. The solid (dashed) lines show the distribution with (without) the normalisation parameters $\lambda^{\mathrm{MB}}$ applied. Note that some of the $\cos\theta_{\mu}$ slices have been combined in this plot for ease of presentation.}
    \label{fig:best_fit_mb}\vspace{-12pt}
  \end{figure}

\begin{table}[h!]
\centering
\footnotesize
{\renewcommand{\arraystretch}{1.2}
\begin{tabular}{ccccccc}
\hline
Fit type & $\chi^{2}$/DOF & $M_{\mathrm{A}}$ (GeV) & MEC (\%) & $p_{\mathrm{F}}$ (MeV) & $\lambda_{\nu}^{\mathrm{MB}}$ & $\lambda_{\bar{\nu}}^{\mathrm{MB}}$ \\
\hline
Rel. RPA & 97.84/228 & 1.15$\pm$0.03 & 27$\pm$12 & 223$\pm$5 & 0.79$\pm$0.03 & 0.78$\pm$0.03 \\
Non-rel. RPA & 117.87/228 & 1.07$\pm$0.03 & 34$\pm$12 & 225$\pm$5 & 0.80$\pm$0.04 & 0.75$\pm$0.03 \\
SF+MEC & 97.46/228 & 1.33$\pm$0.02 & 0 (at limit) & 234$\pm$4 & 0.81$\pm$0.02 & 0.86$\pm$0.02 \\
\hline
\end{tabular}}
\caption{Best fit parameter values for combined fits to all four CCQE datasets simultaneously, for the RFG+RPA+MEC and SF+MEC fits.}\label{tab:best_fit}
\end{table}
The best fit distributions for the SF+MEC and relativistic RFG+RPA+MEC models are compared with data for MINER$\nu$A in Figure~\ref{fig:best_fit_minerva} and MiniBooNE in Figure~\ref{fig:best_fit_mb}. In the legends of these figures, two $\chi^{2}$ values are given: the contribution from that dataset to the total, and the total $\chi^{2}_{min}$ in parentheses. Note that the contribution from each MINER$\nu$A dataset ignores correlations, and $\chi^{2}_{\mathrm{MIN\;total}} \neq \chi^{2}_{\mathrm{MIN}\;\nu} + \chi^{2}_{\mathrm{MIN}\;\bar{\nu}}$, so the values shown in the legends should be treated with some caution. It is clear from Figure~\ref{fig:best_fit_mb} that MiniBooNE is not dominating the fit, as might be expected given the large number of bins in the MiniBooNE datasets. Indeed, the fit exploits the fact that without a covariance matrix for MiniBooNE, $\chi^{2}_{\mathrm{MB}} \approx \chi^{2}_{\mathrm{MIN}}$. It is also clear that neither model fits all of the datasets perfectly, although this is not reflected by the $\chi^{2}_{min}$ values in Table~\ref{tab:best_fit}.
\subsection{Can these results be trusted?}
A provocative, but important question to ask is whether these fit results can be trusted. It has been remarked that the $\chi^{2}_{min}$ values suggest very good agreement for both models, but there is clearly tension with the data, and the best fit parameter values for the SF+MEC model suggest strong tensions between the datasets.

Without the MiniBooNE covariance matrix, the $\chi^{2}$ contributions from MiniBooNE are much lower than the number of degrees of freedom it contributes would suggest, and Gaussian statistics do not work. This results in two problems: standard goodness of fit tests such as the Pearson $\chi^{2}_{min}$ test (SGoF) are unreliable, indicating unrealistically good fits; and $\Delta\chi^2 = 1$ is no longer an appropriate method for calculating parameter errors. We tackle both of these issues with the Parameter Goodness of Fit (PGoF) test.

The PGoF test statistic is defined by Equation~\ref{eq:pgof}~\cite{pgof-2003}:
\begin{align}
  \small
  \chi_{\mbox{\scriptsize{PGoF}}}^{2}(\vec{\mathbf{x}}) &= \chi^{2}_{tot}(\vec{\mathbf{x}}) - \sum^{D}_{r=1} \chi^{2}_{r, \: min}(\vec{\mathbf{x}}), &  P_{\mbox{\scriptsize{PGoF}}} = \sum_{r=1}^{D}P_{r} - P_{tot},
  \label{eq:pgof}
\end{align}
\noindent where $D$ are the number of datasets, $\chi^{2}_{tot}$ is the minimum $\chi^{2}$ in a fit to all $D$ datasets, and $P_r$ and $P_{tot}$ are the number of free parameters varied in each fit.

The PGoF tests the compatibility of different datasets in the framework of the model. Put simply, it tests whether fits to subsets of the data pull the best fit parameter far from those found in fits to the complete dataset, which would indicate tension between the datasets. A low PGoF value implies a poor fit. Although the PGoF test still assumes that the datasets follow a $\chi^{2}$ distribution, it does not matter how many bins (degrees of freedom) each dataset has. By assuming that without the covariance matrix, MiniBooNE data follows a $\chi^{2}$ distribution, but with a lower, effective number of degrees of freedom, we can use the results of the PGoF test.

The PGoF results for various subsets of the data are shown for the relativistic RFG+RPA+MEC (SF+MEC) model in Table~\ref{tab:pgof_rfg} (\ref{tab:pgof_sf}). The $\chi^2_{min}$ value for each subset of the data is found by minimising Equation~\ref{eq:chi2} with only the relevant terms included. In each fit, $p_{\mathrm{F}}$, MEC normalisation, $M_{\mathrm{A}}$ and relevant MiniBooNE normalisation parameters are allowed to vary. An explicit example of how the PGoF statistic is calculated is given in Equation~\ref{eq:pgof_example}, which corresponds to the final row in Tables~\ref{tab:pgof_rfg} and~\ref{tab:pgof_sf}:
\begin{equation}
  \small
  \chi^{2}_{\mathrm{PGoF \: \nu \: vs \: \bar{\nu}}} = \chi^{2}_{\mathrm{MB \: \bar{\nu} \: + \: MN \: \bar{\nu}}} - \chi^{2}_{\mathrm{MB \: \bar{\nu}}} - \chi^{2}_{\mathrm{MN \: \bar{\nu}}}
  \label{eq:pgof_example}
\end{equation}\vspace{-12pt}
\begin{table}[htb]
  \small
  \centering
 {\renewcommand{\arraystretch}{1.2}
  \begin{tabular}{ccccc}
    \hline
    & $\chi^{2}_{min}$/DOF & SGoF (\%) & $\chi^{2}_{\mbox{\scriptsize{PGoF}}}$/DOF & PGoF (\%) \\
    \hline
    All & 97.8/228 & 100.00 & 17.9/6 & 0.66 \\
    MINER$\nu$A ($\nu + \bar{\nu}$) & 23.4/13 & 3.74 & 1.0/3 & 79.03 \\
    MiniBooNE ($\nu + \bar{\nu}$) & 58.6/212 & 100.00 & 2.0/3 & 57.69 \\
    $\nu$ (MB + MIN) & 62.6/142 & 100.00 & 16.1/3 & 0.11 \\
    $\bar{\nu}$ (MB + MIN) & 38.5/83 & 100.00 & 6.1/3 & 10.75 \\
    MINER$\nu$A vs MiniBooNE & 97.8/228 & 100.00 & 15.9/3 & 0.12 \\
    $\nu$ vs $\bar{\nu}$ & 97.8/228 & 100.00 & -3.3/3 & 100.00 \\
    \hline
  \end{tabular}}
  \caption{PGoF statistics for the relativistic RFG+RPA+MEC model. In each fit, $M_{\mathrm{A}}$, $p_{\mathrm{F}}^{\mbox{\scriptsize{RFG}}}$, MEC normalisation and relevant MiniBooNE normalisation parameters are allowed to vary.}
  \label{tab:pgof_rfg}
\end{table}

It is clear that there is a great deal of tension in the SF+MEC model. Although there is some tension in the relativistic RFG+RPA+MEC model, it is clearly a more consistent fit to the data. For this reason, the SF+MEC model was not selected as the default T2K MC model, and is not discussed further in these proceedings. It is reassuring to see good agreement between the neutrino and antineutrino datasets for the RFG+RPA+MEC model, as shown on the final line of Table~\ref{tab:pgof_rfg}, indicating that the same parameters can be used for neutrino and antineutrino T2K analyses.
\begin{table}[htb]
  \small
  \centering
 {\renewcommand{\arraystretch}{1.2}
  \begin{tabular}{ccccc}
    \hline
    & $\chi^{2}_{min}$/DOF & SGoF (\%) & $\chi^{2}_{\mbox{\scriptsize{PGoF}}}$/DOF & PGoF (\%) \\
    \hline
    All & 97.5/228 & 100.00 & 41.1/6 & 0.00 \\
    MINER$\nu$A ($\nu + \bar{\nu}$) & 12.6/13 & 47.75 & 1.0/3 & 79.49 \\
    MiniBooNE ($\nu + \bar{\nu}$) & 50.2/212 & 100.00 & 6.5/3 & 8.92 \\
    $\nu$ (MB + MIN) & 54.8/142 & 100.00 & 25.1/3 & 0.00 \\
    $\bar{\nu}$ (MB + MIN) & 34.1/83 & 100.00 & 8.5/3 & 3.61 \\
    MINER$\nu$A vs MiniBooNE & 97.5/228 & 100.00 & 34.6/3 & 0.00 \\
    $\nu$ vs $\bar{\nu}$ & 97.5/228 & 100.00 & 8.5/3 & 3.59 \\
    \hline
  \end{tabular}}
  \caption{PGoF statistics for the SF+MEC model. In each fit, $M_{\mathrm{A}}$, $p_{\mathrm{F}}^{\mbox{\scriptsize{SF}}}$, MEC normalisation and MiniBooNE normalisation parameters are allowed to vary.}
  \label{tab:pgof_sf}\vspace{-12pt}
\end{table}

MINUIT uses $\Delta\chi^2 = 1$ to define $1\sigma$ parameter errors, which are not appropriate for non-Gaussian datasets, so the errors returned by our fits are suspicious. This problem has also been faced by parton density distribution fitters~\cite{pumplin2000_pdfs}, who overcome it by inflating the value of $\Delta\chi^2$ used to define the errors, although there is no general solution offered for choosing that value. 

The PGoF gives a value of the incompatibility between datasets: how much the $\chi^2$ increases between the best fit points of subsets of the data, and the best fit for the combined dataset. The PGoF value can therefore be used as a measure of how much the errors must be inflated to cover the difference between the best fit parameter values from the combined fit, and the best fit values found in fits to individual experiments. The rescaled error therefore includes our lack of understanding of the differences in the fit results for different subsets of the data. We select $\Delta \chi^{2} = \sqrt{\chi^{2}_{\mbox{\scriptsize{PGoF}}}/\mathrm{DOF}_{\mbox{\scriptsize{PGoF}}}}$. There is some ambiguity over which PGoF value to use, but we use the ``MINER$\nu$A vs MiniBooNE'' row of Table~\ref{tab:pgof_rfg}, as it is the most conservative. This results in a rescaling factor of approximately 2.4.

\subsection{Final fit results}
\begin{figure}[htb]
  \centering
  \includegraphics[width=0.4\textwidth]{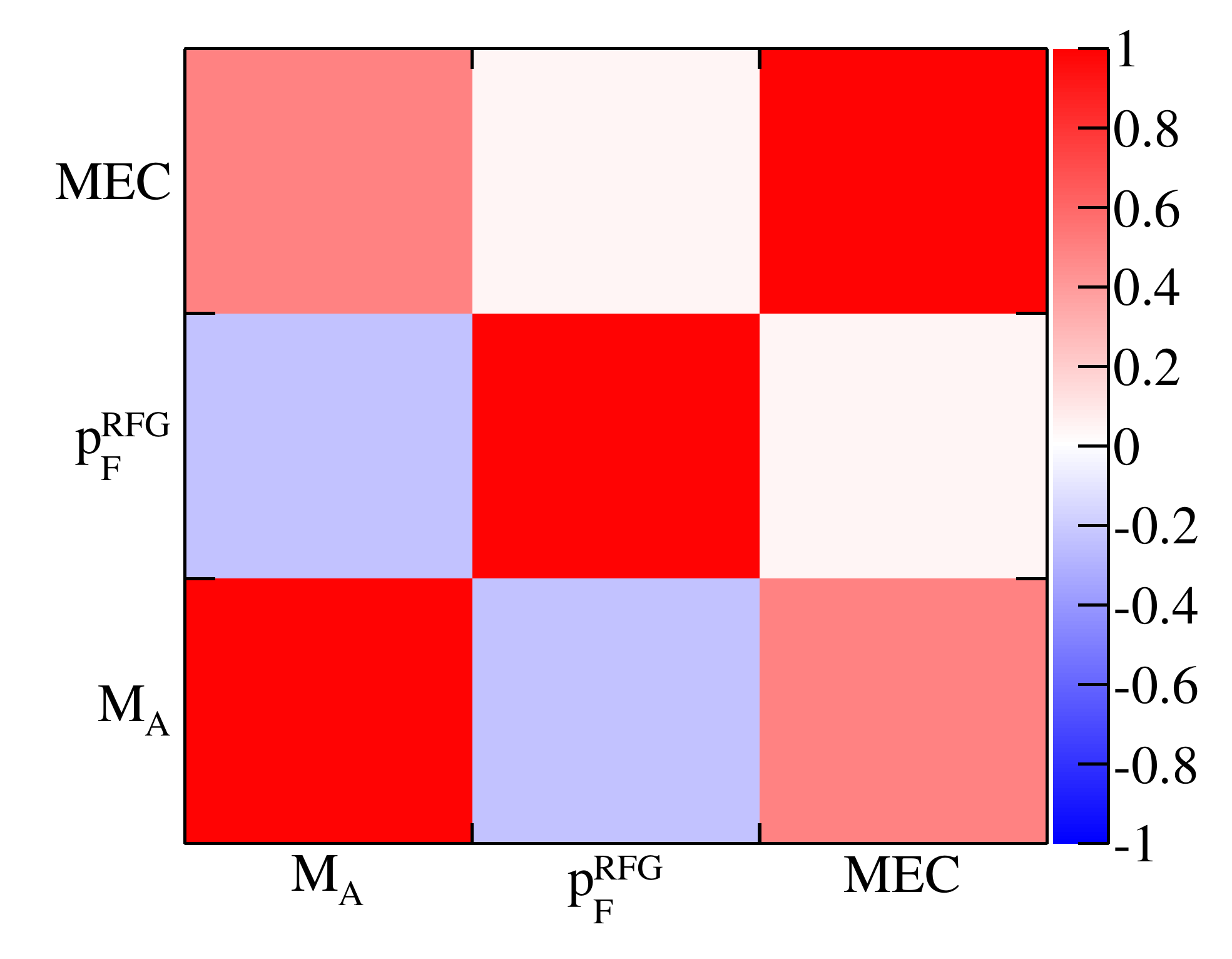}\vspace{-6pt}
  \caption{Correlation matrix for the CCQE parameters for the relativistic RFG+RPA+MEC model.}\vspace{-2pt}
  \label{fig:corr_mat}
\end{figure}

The correlations between the CCQE parameters for the relativistic RFG+RPA+MEC model produced in this work are shown in Table~\ref{fig:corr_mat}. The final errors on the fitted parameters are shown in Table~\ref{tab:final_errors}, where the unscaled errors are also shown for comparison. When an overall normalisation parameter for the CCQE interaction channel was included in the RFG+RPA+MEC fit, there was no tendency to pull it away from unity. The CCQE normalisation was therefore fixed at the nominal, and no error was assigned.
\begin{table}[h!]
\small
\centering        
 {\renewcommand{\arraystretch}{1.2}
  \begin{tabular}{ccccc}
    \hline
    Fit type & $\chi^{2}$/DOF & $M_{\mathrm{A}}$ (GeV) & MEC (\%) & $p_{\mathrm{F}}$ (MeV) \\
    \hline
    Unscaled & \multirow{2}{*}{97.84/228} & 1.15$\pm$0.03 & 27$\pm$12 & 223$\pm$5 \\
    PGoF scaling & & 1.15$\pm$0.06 & 27$\pm$27 & 223$\pm$11 \\
    \hline
\end{tabular}}
  \caption{The final, scaled, errors for the CCQE parameters fitted to the relativistic RFG+RPA+MEC model in this work. The unscaled errors are shown for comparison.}
  \label{tab:final_errors}\vspace{-6pt}
\end{table}
\section{Summary}\label{sec:summary}
A number of new models have been implemented in NEUT ready for inclusion in future T2K analyses. There are currently two candidate CCQE models which could be used as the default MC model: SF+MEC and RFG+RFG+MEC (with two RPA model options). This work describes fits to both models using CCQE data from MINER$\nu$A and MiniBooNE to improve the nominal T2K model. A generic framework was developed to allow new models and new datasets to be included in future iterations of the fit, in order to keep the T2K model up to date. In this iteration, we have selected the relativistic RFG+RPA+MEC model as the default CCQE model, and have obtained a set of parameters which describe all of the data well, although the best fit parameter values suggest that there is some tension between theory and data.

\small
\begin{spacing}{0.89}
\bibliographystyle{JHEP}
\bibliography{NuFact_proceedings}
\end{spacing}
\normalsize
\end{document}